\documentclass[letterpaper,10pt]{article}
\usepackage{biblatex} 
\usepackage{authblk}
\usepackage{subfig}
\usepackage{graphicx}

\usepackage{amsmath, amsfonts, amssymb, amsthm}

\begin{document}

\title{Correcting public opinion trends through Bayesian data assimilation}


\author{Robin Hendrickx}
\author{Rossella Arcucci\thanks{corresponding: r.arcucci@imperial.ac.uk}}
\author{Julio Amador D\'iaz Lopez}
\author{Yi-Ke Guo}
\author{Mark Kennedy}

\affil{Data Science Institute, Imperial College London (UK)}

\date{}

\maketitle

\begin{abstract}
Measuring public opinion is a key focus during democratic elections, enabling candidates to gauge their popularity and alter their campaign strategies accordingly. Traditional survey polling remains the most popular estimation technique, despite its cost and time intensity, measurement errors, lack of real-time capabilities and lagged representation of public opinion. In recent years, Twitter opinion mining has attempted to combat these issues. Despite achieving promising results, it experiences its own set of shortcomings such as an unrepresentative sample population and a lack of long term stability. This paper aims to merge data from both these techniques using Bayesian data assimilation to arrive at a more accurate estimate of true public opinion for the Brexit referendum. This paper demonstrates the effectiveness of the proposed approach using Twitter opinion data and survey data from trusted pollsters. Firstly, the possible existence of a time gap of 16 days between the two data sets is identified. This gap is subsequently incorporated into a proposed assimilation architecture. This method was found to adequately incorporate information from both sources and measure a strong upward trend in Leave support leading up to the Brexit referendum. The proposed technique provides useful estimates of true opinion, which is essential to future opinion measurement and forecasting research.
\end{abstract}




\section{Introduction}
Traditional opinion polling is a well-established technique with a proven ability to produce relatively accurate results. However, its poor scalability makes it both cost and time-intensive. This restricts the ability to conduct polls in real-time or on adequately large samples. In addition, the time taken to perform and compile surveys causes polls to represent a lagged mood of the electorate \cite{RefWorks:doc:bovet}\cite{RefWorks:doc:Brendan}.\newline

To counter the shortcomings of polls, researchers have been flocking towards methods that measure opinion through social media. Their advantages are clear: they potentially allow for a highly scalable and hence cost-effective method of analysing public mood instantaneously. However, most research has been criticised for its 'ad-hoc' nature  \cite{RefWorks:doc:Gayo2011}\cite{RefWorks:doc:Gayo2012}. Despite its drawbacks, it has been argued that traditional survey polls remain a more accurate opinion predictor \cite{RefWorks:doc:Gayo2012}. However, models incorporating both twitter and polling data have shown improved performance over models based solely on polls \cite{twitterAndPolls}. \newline

Despite having made significant strides, most research has focused on forecasting polls, failing to recognise that polls in themselves are an inaccurate representation of public opinion. This paper proposes a Bayesian data assimilation method which arrives at an improved estimate of public opinion, providing a more reliable ground truth for future public opinion research. Although this research focuses on the Brexit elections, the techniques used can be applied to a broad array of data assimilation issues that attempt to assimilate time series from different measurement sources.\\

The main contributions of this project are the calculation of a time delay between Twitter opinion and polls, and the introduction and validation of a custom data assimilation architecture which estimates true public opinion.

\begin{enumerate}
  \item The existence of a time delay between Twitter opinion and polling measurements is postulated through the use of a time series fitting approach. The procedure involved re-scaling and smoothing the time series, and measuring the parameters which minimise the RMSE. The time delay using this method was found to be 16 days. It was estimated that Twitter opinion precedes true sentiment and polls lag true sentiment by 14 and 2 days respectively. 
  \newline
  \item The proposed assimilation architecture applies the discovered time shift to both Twitter and polling data and merges the resulting sets optimally through Optimal Interpolation (OI). This method efficiently incorporates information from both sources and benefits from algorithmic simplicity and computational efficiency. It also recognised a strong growth in support for the Leave campaign leading up to the referendum. 
\end{enumerate}

The rest of the paper is structured as follows.
In the second section, background is provided on Twitter opinion measurement techniques and their challenges. The third section describes the data sets used in this work and evaluates the existence of a time gap between Twitter and polling data. The fourth section evaluates and discusses the proposed data assimilation scheme. The fifth and sixth sections show the testing of the proposed model, the set-up of hyperparameters and the results of the assimilation process.
The seventh and final section concludes the paper and introduces future works.


\section{Background and related work}
Despite being the most accurate measurement technique to date, traditional polling still suffers from some drawbacks. Firstly, the methods used suffer from a high financial cost and are inherently time intensive. Even with market research companies moving some polls online, they still require screening and active participation of the respondents. As a result, sample sizes are relatively small. Additionally, there are demographic and geographic biases present in the polled sample which could skew results. Surveys are also facing a steep decline in response rates, going from 36\% in 1997 to only 9\% in 2016 \cite{RefWorks:doc:pewResearch}. In contrast, Twitter usage shows strong growth: around 25\% of the UK population used the platform in 2018 to a certain extent \cite{RefWorks:doc:Statista}.\newline

An expanding body of research has focused on predicting election results employing a range of different methods. The simplest approach consists of counting the number of Tweets referring to a candidate, by using either keywords in the tweet corpus or 'hashtags' that mention a specific candidate or camp. Note that hashtags are topical keywords that users include in their Tweets. They serve as topic labels on Twitter, allowing users to quickly search for Tweets containing the same hashtag. As an example, '\#StrongerIn' and '\#TakeControl' were two important hashtags that recurred in Tweets for the Remain and Leave campaign \cite{lopez2017predicting}. Whilst this simple approach yielded reasonable accuracy in some cases \cite{RefWorks:doc:Tumasjan2010}\cite{RefWorks:doc:Tumasjan2011}, the main critique is that attention towards a certain candidate does not necessarily translate into political support \cite{RefWorks:doc:Gayo2011}\cite{RefWorks:doc:Gayo2012}. A more advanced approach consists of performing a sentiment analysis using simple lexicons or supervised learning on Tweets which mention the specific topic. This method generally outperforms simple counting, albeit with some exceptions, but does not consistently predict polling results. Bassilakis et al.\,\cite{bassilakis2018converging} demonstrated this approach by assessing post-Brexit sentiment through training a Naive Bayes classifier to memorise features of Tweets of prominent Leave and Remain accounts. Amador et al.\,\cite{lopez2017predicting} efficiently employed a Support Machine Vector to learn features of both Remain and Leave-supporting Tweets. This Tweet-level method was benchmarked against a user-level approach, using a multinomial Naive Bayes classifier to infer the political preference of each user. Such user-level approaches have gained traction primarily because Tweets per user are highly skewed: in the US, 80\% of Tweets come from the 20\% most active users \cite{RefWorks:doc:pewResearch2}. Additionally, Celli et al.\,\cite{RefWorks:doc:Celli} successfully showed that language-independent stylometric features such as emoticons and special characters are able to classify Brexit Tweets to a high degree of accuracy.\newline

Other research has focused on incorporating graph theory, using 'retweets', where a user forwards a copy of another Tweet,  and 'replies', where a user reacts to another Tweet, to build connectivity networks. Becatti et al.\, \cite{becatti2019extracting} leveraged retweets to identify the formation of political alliance networks and viral content during the Italian 2018 elections. Bovet et al.\,\cite{RefWorks:doc:bovet} incorporated graph theory to build both a co-occurrence network of hashtags and a user connectivity graph for the 2016 US presidential campaign. For the hashtag network, an edge was drawn when the co-occurrence between hashtags was significant. This led to the interesting finding that hashtags separated into two distinct clusters, in support of Clinton and Trump respectively. This allowed them to efficiently build a training set, on which they subsequently trained different types of classifiers which were able to closely predict polling results. Furthermore, they investigated the existence of such a distinct split on the topic of climate change and in a multiparty environment, and discovered similar separated clusters in both cases. It is highly plausible such a split exists on the Brexit topic as well, which corroborates the use of hashtags to build training sets as used in Amador et al.\,\cite{lopez2017predicting}. Finally, Bovet et al.\,\cite{RefWorks:doc:bovet} divided users into three different components according to the levels of interactivity. Interestingly, they found that whilst the entire network was Clinton dominated, the most interactive component was dominated by Trump supporters.\newline

\begin{figure}[h!]
\centering
\includegraphics[width=8cm]{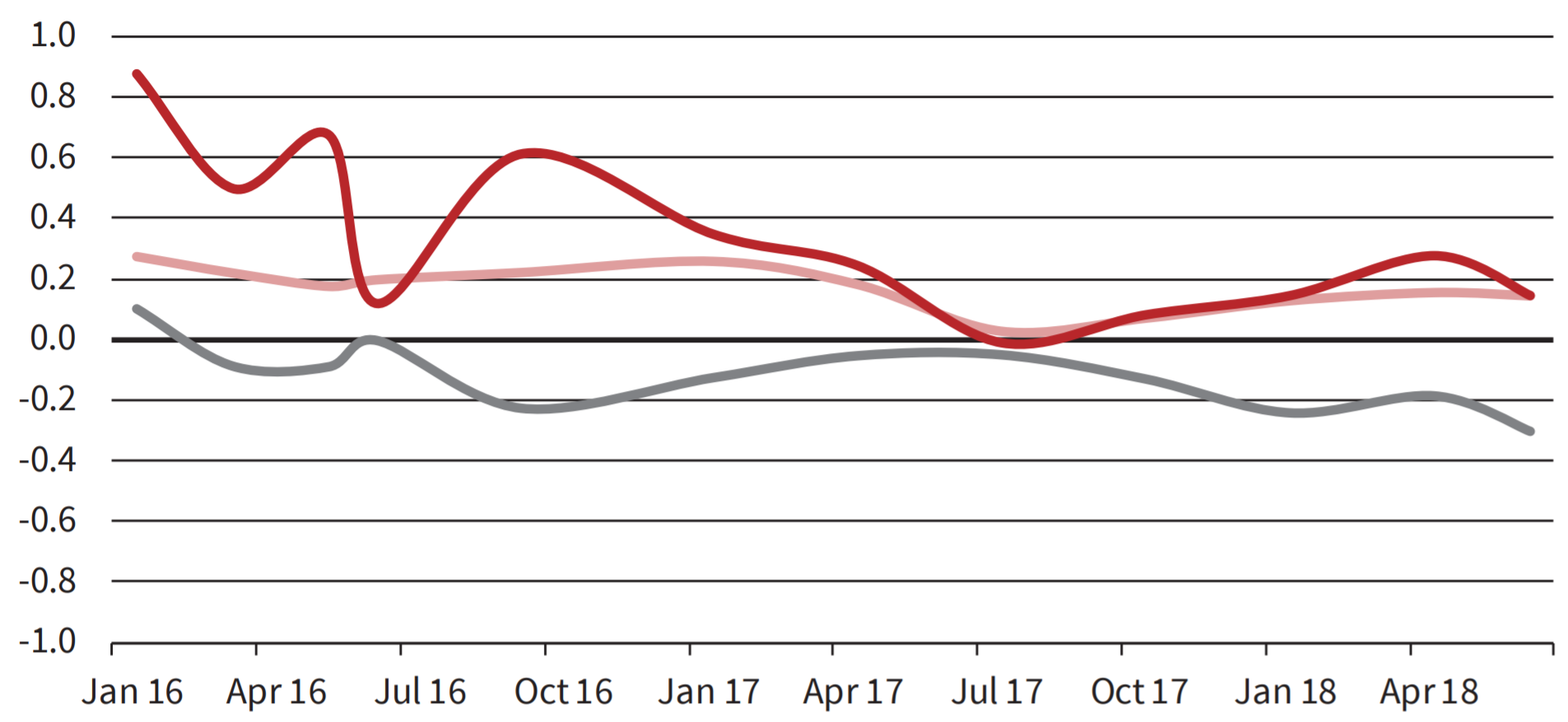}
\centering
\caption{Classification of post-referendum Brexit sentiment of the Conservative party (grey), Labour party (pink) and Liberal Democrats (red) through the use of a Naive Bayes classifier trained on significant lexicological features \cite{bassilakis2018converging}.}
\label{fig:bassilakis}
\end{figure}

\section{Data}
Twitter data used in this project was acquired by Amador et al.\ \cite{lopez2017predicting}:\ 
using the Twitter 'firehose' API they harvested over 30M Tweets by querying 'Brexit' as a generic search term, Brexit related hashtags and important Brexit-related accounts. After preprocessing, they built a training set using hashtags that indicate support for both the Leave and Remain camp. Bovet et al.\ \cite{RefWorks:doc:bovet} used a similar approach to build their training set, and demonstrated its validity by creating a co-occurrence network of hashtags. This resulted in two main clusters, with one set of supporting Trump and the other Clinton. It is highly plausible such a strong grouping exists for the harvested Brexit Tweets as well. 
Amador et al.\,\cite{lopez2017predicting} subsequently trained an SVM classifier and classified over 23M Tweets, assigning them to either Leave or Remain when the probability of the Tweet showing support for either camp was higher than 70\%. This resulted in 182,553 Tweets supporting Leave and 310,932 Tweets supporting Remain. The Tweets range from the 7th of January to the 30th of June 2016. \\

In this paper, we decided to discard data collected prior to the 1st of March to leave out days with an insignificant amount of Tweets. The Tweets are processed using an aggregation function $\mathcal{A}$, which sums the amount of Tweets per day for each camp. The result is then divided by the total number of Tweets that day to arrive at a daily support percentage (normalisation function $\mathcal{G}$). The resulting values are here denoted as $x_k$, where $k$ represents the temporal step (day):

\begin{equation}\label{dynamicEq}
    x_k = \mathcal{G} ( \mathcal{A} (SVM (\textrm{Tweets}))) + w_k
\end{equation}
where $w_k$ is the process noise, drawn for a multivariate normal distribution with zero mean. 
The values $x_k$ will be used as ``prior'' estimations of public opinion in the model we present in the next section.


\begin{figure}[h!]
\centering
\includegraphics[width=12cm]{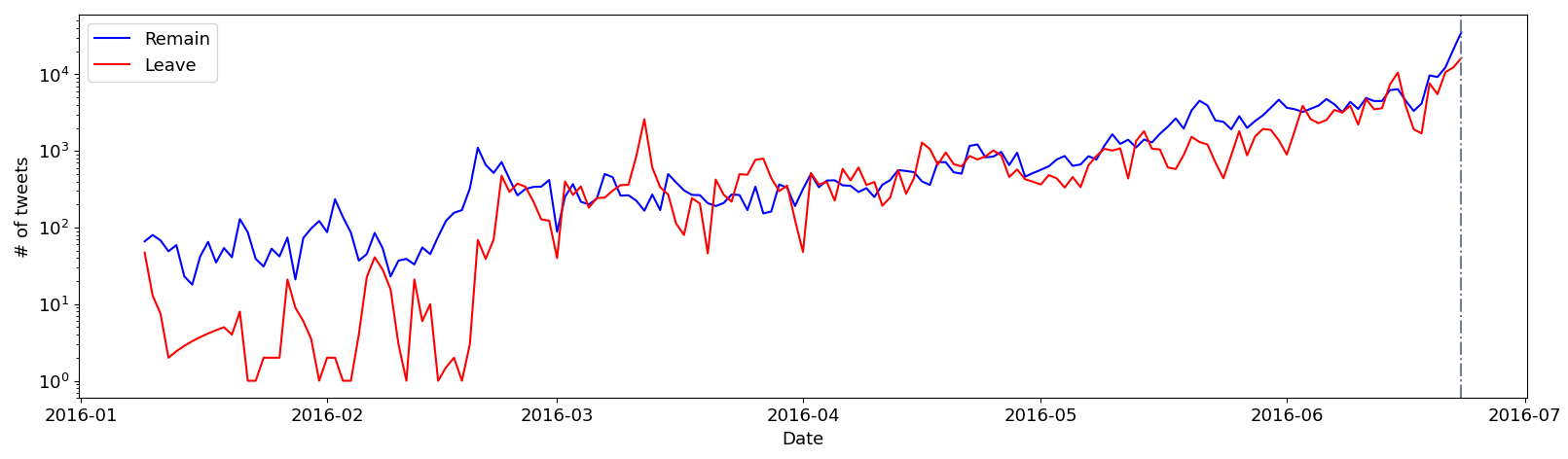}
\centering
\caption{Tweets classified as supporting Leave and Remain by Amador et al.\ \cite{lopez2017predicting} aggregated by day.}
\label{fig:tweets}
\end{figure}

\begin{figure}[h!]
\centering
\includegraphics[width = 12cm]{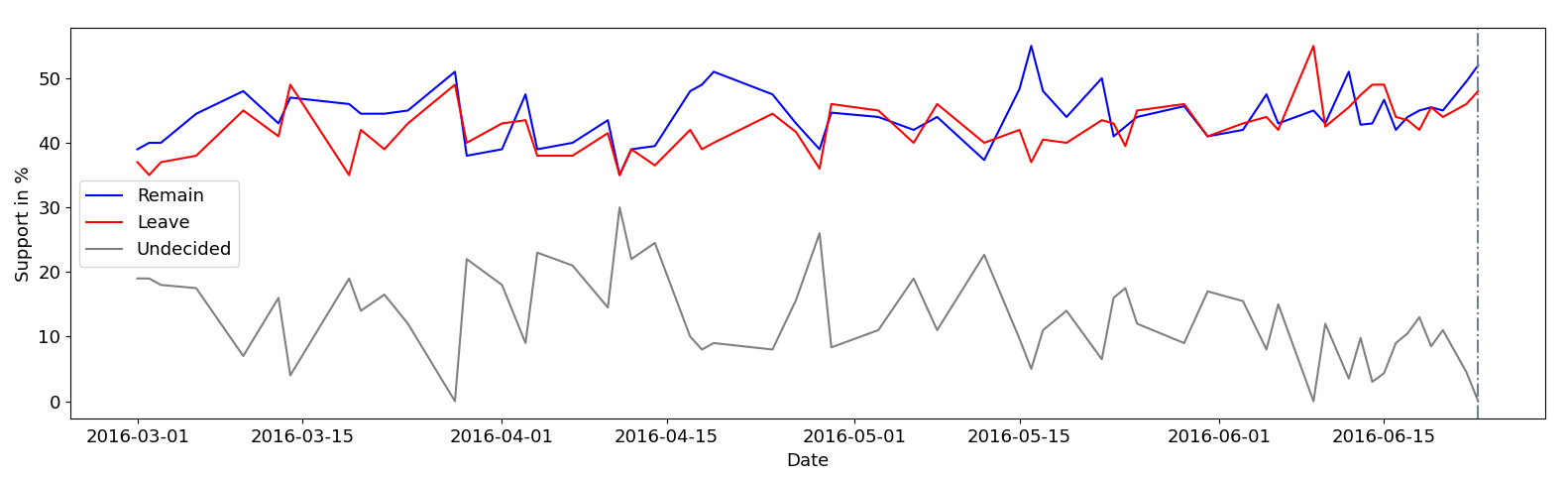}
\centering
\caption{Raw polling data normalised by day.}
\label{fig:polls}
\end{figure}

A total of 97 polls were collected which were conducted from the 1st of March to the 23rd of June 2016. Of those polls, 64 were conducted online whilst the remainder consists of telephone polls. Every included poll was carried out by members of the British Polling Council, a consortium of trusted pollsters who abide by specific transparency standards. A significant portion of the data was scraped from the UK Polling Report \cite{RefWorks:doc:pollingreport}, whilst some surveys were sourced directly from relevant pollsters. Figure \ref{fig:polls} shows the evolution of the support for Remain, Leaven and the share of undecided voters over time.\newline

As a preliminary step, the correlation between both the Twitter and polling time series was investigated using the Pearson product-moment correlation. The calculation was performed in two different ways since both polls and Twitter time series contain missing values. The first approach is a "complete case" analysis: when a value is missing in either data set, the pair is omitted. In the second approach, missing data points were linearly interpolated before calculating the correlation. Note that correlation results after interpolation potentially grow spurious for sparse data \cite{hassani2019evaluating}. The formula used is shown in (\ref{eq:corr}).

\begin{equation}
\label{eq:corr}
\rho = \frac{cov(X,Y)}{\sigma_x \sigma_y}
\end{equation}
\newline

Results were insignificant: -0.03 for the complete case analysis, and -0.04 for the case where interpolation was used. The poor correlation was expected since a time lag likely exists between polling data and Twitter sentiment: recent state-of-the-art research linking polls to twitter data has estimated this lag to be around 10 days for the 2016 US presidential campaign \cite{RefWorks:doc:bovet}.
Two main reasons exist for the measured lag: Firstly, polls are conducted over a period of multiple days.  This creates a misalignment between the official release date of a poll and the day on which the sentiment was actually measured [58]. Secondly, it is conceivable that instantaneous Twitter opinion precedes informed decision making with regard to elections by a few days.\newline

To estimate this postulated lag, four different experiments were performed which attempt to fit the Twitter to the polling series. Firstly, the Twitter data was smoothed using Locally Weighted Scatterplot Smoothing (LOWESS), a method chosen because of its favourable handling of outliers, peaks and troughs \cite{MAdownsides2}. In the first experiment, the smoothed Twitter series was shifted forward in time by $t_d$ days in order to minimise the Root Mean Squared Error (RMSE). 
The second experiment extends upon the first by rescaling the smoothing Twitter series, using equation (\ref{eq:corrLaglowess}). 

\begin{equation}
    r^{'}_w(i) = Al(r_w(i-t_d)) + b
    \label{eq:corrLaglowess}
\end{equation}

Note that $l(.)$ indicates the LOWESS smoothing algorithm, $r(i)$ the ratio of supporters at day $i$ and $t_d$ the timeshift in days. The remaining terms $A$ and $b$ are rescaling parameters. These parameters were once again optimised through minimisation of the RMSE between the two series, for which the formula is shown in \ref{eq:RMSE}. 

\begin{equation}
\label{eq:RMSE}
  RMSE = \sqrt{\frac{1}{n}\Sigma_{i=1}^{n}{\Big(\frac{d_i -f}{\sigma}\Big)^2}}
\end{equation}

The symbol $d_i$ denotes a data point of the series, and $f$ the mean of that series. The final two experiments are the same as the first one, with the only difference being that the correlation was taken as the optimisation criterion. The latter two experiments mostly operate as a validation method for the first two, since correlation should be treated with caution with smoothed series due to the possibility of spurious results \cite{hassani2019evaluating}.\newline

\begin{table}[h!]
\centering
\begin{tabular}{l|cccc} 
\hline
\multicolumn{1}{c|}{\textbf{Parameter}} & \textbf{S \& R} & \textbf{S \& R} & \textbf{S} & \textbf{S} \\ 
\hline
$t_d$ & 16 & 16 & 15 & 16 \\
$A$ & 0.45 & 0.60 (0.45) & 1 & 1 \\
$b$ & 29 (26) & 30 (15) & 0 & 0 \\
RMSE & 3.59 & 6.83 & 6.83 & 6.96 \\
correlation & 0.25 & 0.25 & 0.21 & 0.25 \\
optimisation & RMSE & correlation & RMSE & correlation \\
\hline
\end{tabular}
\caption{Optimised values for Remain. Leave results that differ from Remain are mentioned in brackets. Note that S stands for smoothing, and S \& R for smoothing and rescaling. The LOWESS smoothing window used in each case was 15\% of the data.}%
  \label{Tb:fitting}
\end{table}

All four experiments were performed from the 1st of March until the 1st of June. This end date was chosen so the lag could be varied between 0 and 23 days. Note that the window size of the LOWESS procedure was fixed at 15\% of the data, which corresponds to moderate smoothing \cite{MAdownsides2}. Table \ref{Tb:fitting} shows the results for all cases. Each of the experiments seems to point towards the existence of a lag of around 16 days. 
This result is in line with Bovet et al.\ \cite{RefWorks:doc:bovet}.\newline

Note that the correlation is 0.25 or slightly less for both Remain and Leave, which is a considerable improvement upon the correlation of the raw Twitter and polling series but still a relatively weak result. Bovet et al.\,\cite{RefWorks:doc:bovet} managed to obtain higher correlations through their RMSE fitting approach. This divergence can be attributed to: firstly, fewer data points are available for both series due to a lower Twitter adaptation and a smaller number of polling agencies in the UK. Furthermore, traditional polling practices are generally considered more established in the US \cite{hillygus2011evolution}, leading to higher quality polling data. Furthermore, pollsters were unfamiliar with Brexit-type referendums, potentially increasing measurement errors further \cite{RefWorks:doc:Celli}. Taking these considerations into account, and observing that all four experiments converge on a timeshift estimate similar to the one observed in relevant literature, it is reasonable to assume a time lag of 16 days between the two series for the remainder of this paper.\newline

Through examining the polling data, named $z_k$, it was found that the average time taken between start of the fieldwork and release date of the poll is three days for this particular data set. From this it   follows that from the 16 day lag measured between Twitter and polls, a significant portion is due to Twitter preceding public opinion. Analysis of polling durations showed over 87\% took no more than four days whilst only three polls took longer than eight days. In this research, it is assumed that compiling polling results takes one full day on average for the entire data set. Since the average duration between the start and end of polls is three days, it follows that polls need to be shifted backwards by two days to align with the middle of the fieldwork period. This way, polling values line up with the dates of which they reflect public sentiment. Twitter data hence precedes true sentiment by 14 days.
Then, in this paper we assume:
\begin{equation}\label{measEq}
    z_k = H x_k + v_k
\end{equation}
where $H$ is an observation function which shifts the time window of a fixed number of time steps and $v_k$ is the observation noise characterised by a zero-mean Gaussian so that
\begin{equation}\label{eq:observationError}
  v_k\sim\mathcal{N}(0,R_k).  
\end{equation}

In the next section, we present a Bayesian DA model, namely an Optimal Interpolation, to merge twitter data, equation~\eqref{dynamicEq}, with pooling data from equation~\eqref{measEq}, assuming that the distribution of the errors on the data are both zero-mean Gaussian \cite{asch2016data}.

\section{Model}
Bayesian DA, in its most general form, tries to answer questions such as "what can be said about the value of an unknown variable $x_k$ that represents the state of a system, if we have some other measured data $z_k$?". This is the Bayesian context, where we seek a quantification of the uncertainty in our knowledge of the parameters that, according to Bayes’ rule, and for all $k$, it takes the form

\begin{equation}
  p\left( x_k|z_k \right)  = \frac{p\left( z_k|x_k \right) p\left( x_k\right)}{p\left( z_k \right)}
\end{equation}

\noindent Here, $p\left(z_k | x_k\right)$ represents the conditional probability (also known as the likelihood), and $p\left(x_k\right)$ denotes the prior knowledge of the system. The denominator is considered as a normalising factor and represents the total probability of $z_k$. 
There are many DA methods deriving from this formulation \cite{asch2016data}, 
the model we propose in this paper uses Optimal Interpolation (OI), a simplified variant of the vanilla Kalman Filter used to merge data in fixed time steps. 
The model provides an algorithmic solution for a process 
which merges values of a variable $x_k$ with external information $z_k$, also called observations.



\begin{figure}[h!]
\centering
\includegraphics[width = 8cm]{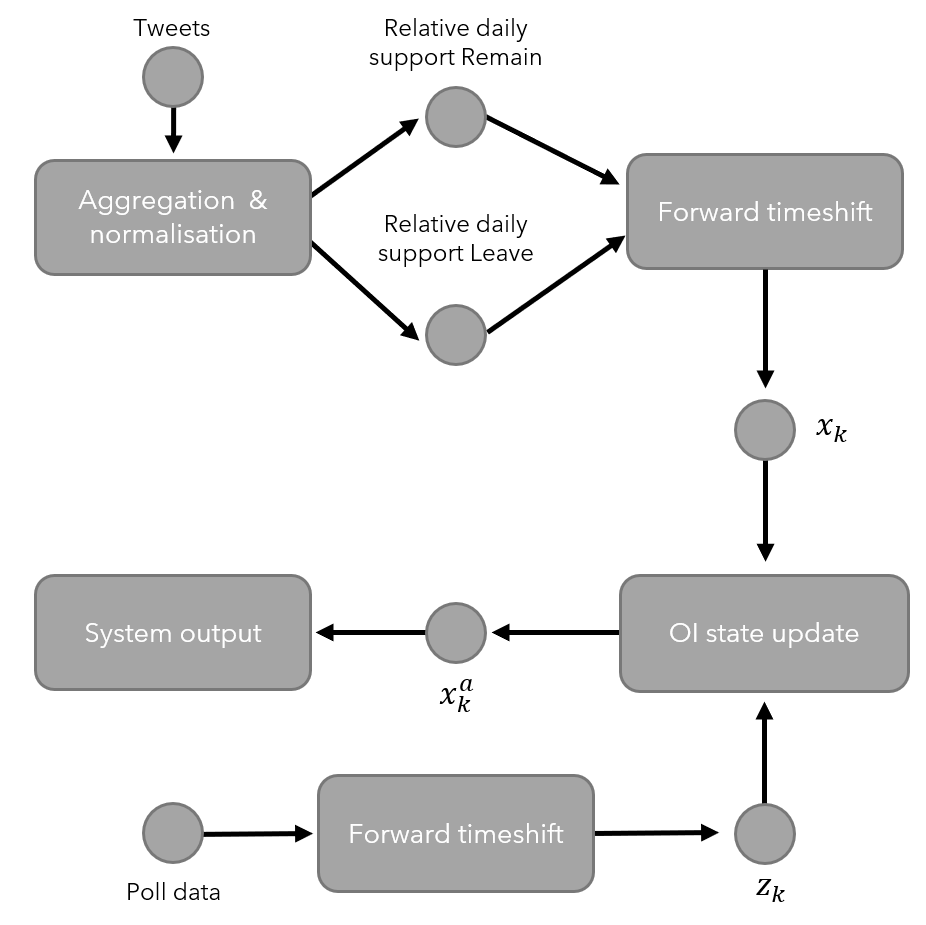}
\centering
\caption{Data assimilation architecture at each time step $k$.}
\label{fig:arch2}
\end{figure}

This paper proposes the following 
assimilation scheme: the Twitter data is directly used as the a priori state estimate, polling data as the system observations, fused through Optimal Interpolation. Since the difference of 16 days between both time series was found to be due to polls lagging behind by 2 days and Twitter preceding public opinion by 14 days, both series are shifted accordingly. Figure \ref{fig:arch2} visualises the architecture used in this project.

The OI model is described in equations~\eqref{kGain}-\eqref{errorUp}. 
The first step of the OI algorithm consist in computing the so called  Kalman gain (\ref{kGain}), then the estimate of the state is updated using both the prediction and the measurement (\ref{stateUp}) and the error covariance matrix is updated (\ref{errorUp}). A full derivation of these equations is available in \cite{asch2016data} or \cite{evensen2009data}.


\begin{equation}
K_{k} = P_{k}^{pr}H^T_k(H P_{k}^{pr}H^T+R_k)^{-1}
\label{kGain}
\end{equation}
\begin{equation}
\label{stateUp}
x_{k}^{a} = x_{k} + K_{k}(z_{k}-Hx_{k})
\end{equation}
\begin{equation}
\label{errorUp}
P_k^a = (I - K_{k}H)P^{pr}_{k}
\end{equation}

\noindent where the superscript $pr$ stands for priori estimate, whilst the superscript $a$ denotes the posterior nature of the estimate. $R_k$ denotes the observation error covariance matrix in \eqref{eq:observationError}, $K_k$ denotes the Kalman gain matrix, determined by minimising the posterior error covariance of which the solution is shown in (\ref{kGain}). When calculating the posterior state estimate in (\ref{stateUp}), the Kalman gain determines the relative importance of the priori estimate (twitter data) and the additional observed values (pooling data): when $R_k$ tends towards zero, the additional observed data is trusted more whilst when $P_k^{pr}$ tends towards zero, the additional observed data is trusted less in favour of the priori estimate $H x_k$.
 $P_k$ represents the error covariance matrix, calculated from the prior and posterior estimate errors $e_k^{pr}$ and $e_k^a$ as shown in (\ref{errorPost}) and (\ref{errorPrior}), where superscript $t$ denotes the true state.

\begin{equation}
P_k^{pr} = cov(e_k^{pr}) = E[e_k^{pr}(e_k^{pr})^T] 
\label{errorPrior}
\end{equation}
with
\begin{equation}
    e_k^{pr} = x_k^{pr} - x_k^t
\end{equation}
where $x_k^t$ denotes the ground truth. 

\begin{equation}
\label{errorPost}
P_k^a = cov(e_k^a) = E[e_k^a(e_k^a)^T] 
\end{equation}
with
\begin{equation}
    e_k^a = x_k^a - x_k^t.
\end{equation}

The covariance matrices, as described in \eqref{errorPrior} and \eqref{errorPost}, needs the knowledge or an estimation of a ground truth which is often not available. Also, the computation involves the multiplications of big matrices which is computationally expensive in some applications involving Big Data.
In order to make the process computationally less expensive, the Optimal Interpolation uses a frozen covariance matrix, which means that $P_k^{pr}$ is fixed for each time step $k$, i.e. $P_k^{pr}=P^b$, $\forall k$, as we will better show in the next section. \\



\section{Experiment}

Experiments were conducted to assess whether the model improves the public opinion prediction for Brexit.
Two different preprocessing steps were applied to the data set before running the assimilation scheme experiments. Firstly, the Twitter data and polling data was shifted in time to account for the measured time gaps. As a second preprocessing step, gaps in polling and Twitter data were linearly interpolated in order to provide a continuous daily time series. This step significantly reduces algorithm complexity with relatively minor impact on result.\newline

\subsection{Hyperparameter estimation}

Central to the implementation of KF-based assimilation schemes is correct parameter and initial state estimation. Misestimation can lead to filter divergence. According to Asch et al.\,\cite{asch2016data}, filter divergence can be diagnosed via multiple criteria. A first symptom is that state error variances are small. Additionally, residual errors becoming biased is a clear warning signal since this violates the white noise assumption of the dynamic model. A final characteristic of filter divergence is a Kalman gain that tends to zero as time increases. Asch et al.\,\cite{asch2016data} additionally propose some empirical guidelines to avoid divergence, two of which are relevant to this research. Firstly, model errors should rather be overestimated than underestimated. Secondly, a lower bound should be used for Kalman gains when possible.\newline

In the case of OI, parameters $R_k$ for the measurement model and and $P^b$ for the frozen covariance matrix need to be identified. Their estimation is crucial to a properly working assimilation scheme: when $R_k$ is overestimated or underestimated, either too little or too much confidence is placed in observed measurements. Similarly for $P^b$, a value that lies either considerably higher or lower than its real value will either mistrust or be overconfident in predicted values.\newline

One way to determine $P^b$, as proposed by Asch et al.\ \cite{asch2016data}, is to use a collection of state vector snapshots $N_e$ from a model free run. Using the first and second statistical moments, the covariance matrix can be calculated (\ref{eq:statCov}). One possibility is to use the prior state estimates as the state vector snapshots as will be experimented with in this paper.
\begin{equation}
P^{b} = \frac{1}{N_e-1}\sum^{N_e}_{l=1}(x_l-x^b)(x_l-x^b)^T 
\label{eq:statCov}
\end{equation}
with
\begin{equation}
    x^b = \frac{1}{N_e}\sum_{l=1}^{N_e}x_l
\end{equation}

In typical KF applications, which use physical measurement equipment such as trajectory estimation through odometers and accelerometers, $R_k$ is provided by the manufacturers. This equipment undergoes rigorous experimentation against ground truth measurements to obtain an optimal estimate \cite{randqmanuf}. When no known $R_k$ values are given, a viable estimation strategy is to measure the system whilst keeping its output constant. After accounting for the mean, the resulting measurements will only contain noise, which allows for the computation of a reasonable estimate. However, it is not always possible to conduct an experiment in which the underlying process is kept constant. Furthermore, sensor variance may differ depending on whether the system is kept constant or changes frequently \cite{chui1989kalman}. \newline

In most cases, parameters $R_k$ is turned into an engineering factor and fine tuned manually \cite{Burat}. Other methods exist that estimate $R_k$ through automated optimisation schemes such as in Chen and Dunnigan \cite{Chen} or Laamari et al.\,\cite{Laamari}, where genetic and evolutionary algorithms were employed. Abbeel et al.\,\cite{Abbeel} proposed four statistical methods which estimate $R_k$ for an extended KF through maximising its predictive performance. They showed that their automatic tuning algorithms significantly outperformed manual tuning. However, both manual and automatic fitting of $R_k$ require knowledge of the ground truth or control inputs, which are both unavailable in the case of public opinion. Therefore, this work will assume different values of $R_k$ and $P^b$ during experimentation and discuss the respective results. Finally, note that schemes exist that tune model errors adaptively as in Solonen et al.\:\cite{Solonen} which will not be investigated in this work. \newline

Logical bounds for the parameters were determined in the following fashion: firstly, to determine $P^b$, an upper and lower bound is estimated through plugging in both the Twitter data and polling data using equation (\ref{measEq}). This results in $266.06\%^{2}$ and $9.85\%^{2}$ respectively.
An estimate for $R_k$ was found by keeping the output of the system constant and calculating the measurement variance. It is assumed sentiment is constant on any one given day, allowing $R_k$ to be estimated by calculating the variance between polls released on the same day. This resulted in $R_k=13.08\%^2$ for both Leave and Remain. The initial state estimate $x_0$ was taken from the shifted Twitter series. Finally, the measurement matrix $H$ was set to one, since measurements and predictions are assumed to be in the same domain.

\section{Assimilation results}

\begin{figure}[h!]
\centering
    \subfloat[$K_k = 0.95$ with $R_k = 13.08\%^2$ and $P^b=266.01\%^2$]{{\includegraphics[width=12cm]{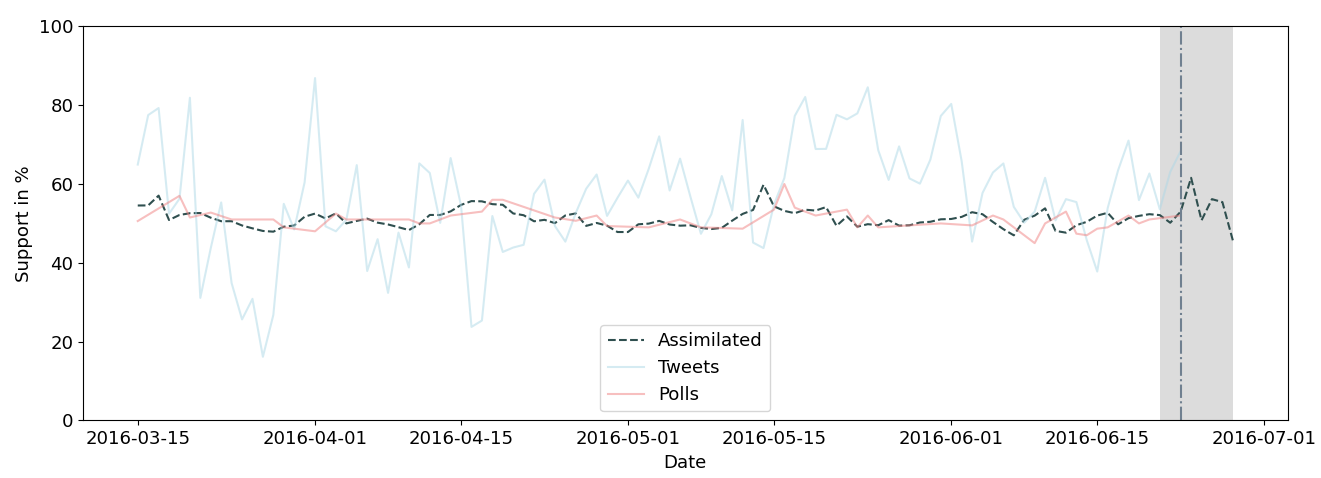} }}%
    \hfill
    \subfloat[$K_k = 0.75$ with $R_k = 13.08\%^2$ and $P^b=45\%^2$ ]{{\includegraphics[width=12cm]{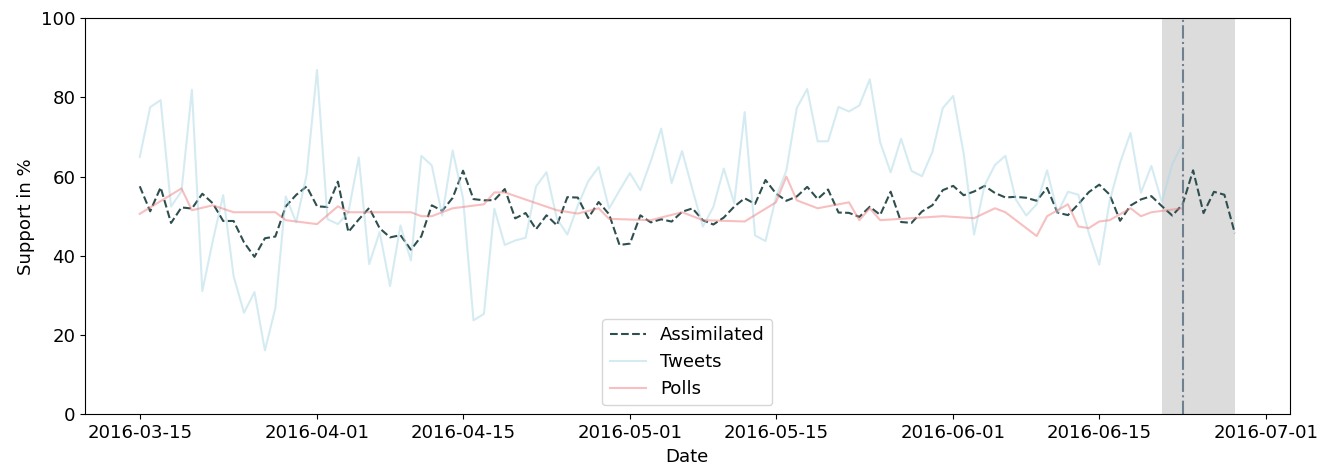} }}%
    \hfill
    \subfloat[$K_k = 0.43$ with $R_k = 13.08\%^2$ and $P^b=10\%^2$]{{\includegraphics[width=12cm]{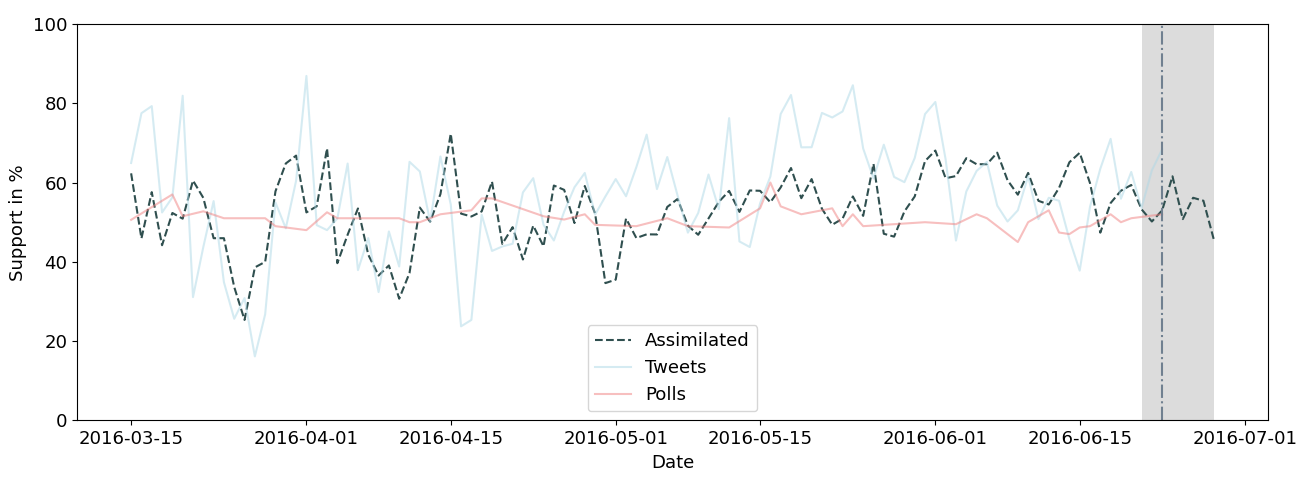} }}%
    \caption{Results for Remain with varying values of $K_k$. Note that the Twitter data plotted is unshifted.}%
    \label{fig:nomodremain}%
\end{figure}

\begin{figure}[h!]
    \centering
    \subfloat[$K_k = 0.95$ with $R_k = 13.08\%^2$ and $P^b=266.01\%^2$]{{\includegraphics[width=12cm]{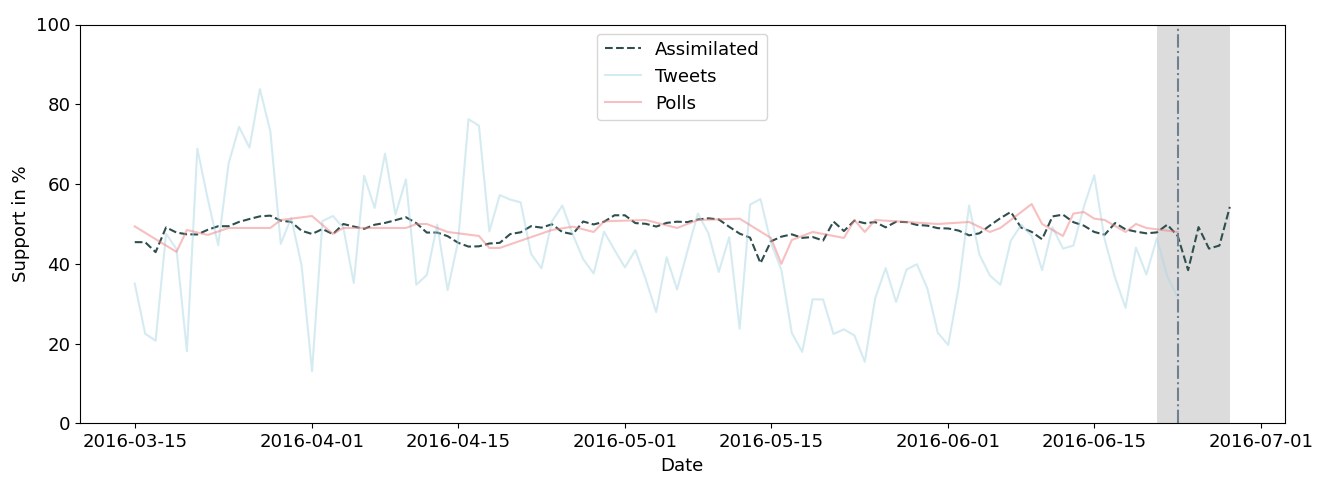} }}%
    \hfill
    \subfloat[$K_k = 0.75$ with $R_k = 13.08\%^2$ and $P^b=45\%^2$ ]{{\includegraphics[width=12cm]{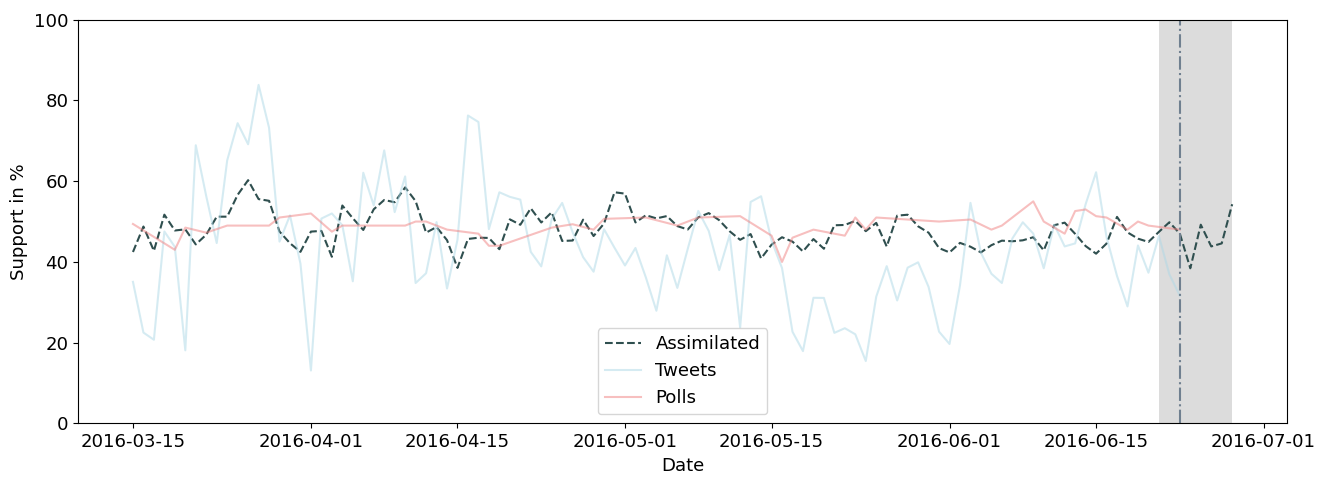} }}%
    \hfill
    \subfloat[$K_k = 0.43$ with $R_k = 13.08\%^2$ and $P^b=10\%^2$]{{\includegraphics[width=12cm]{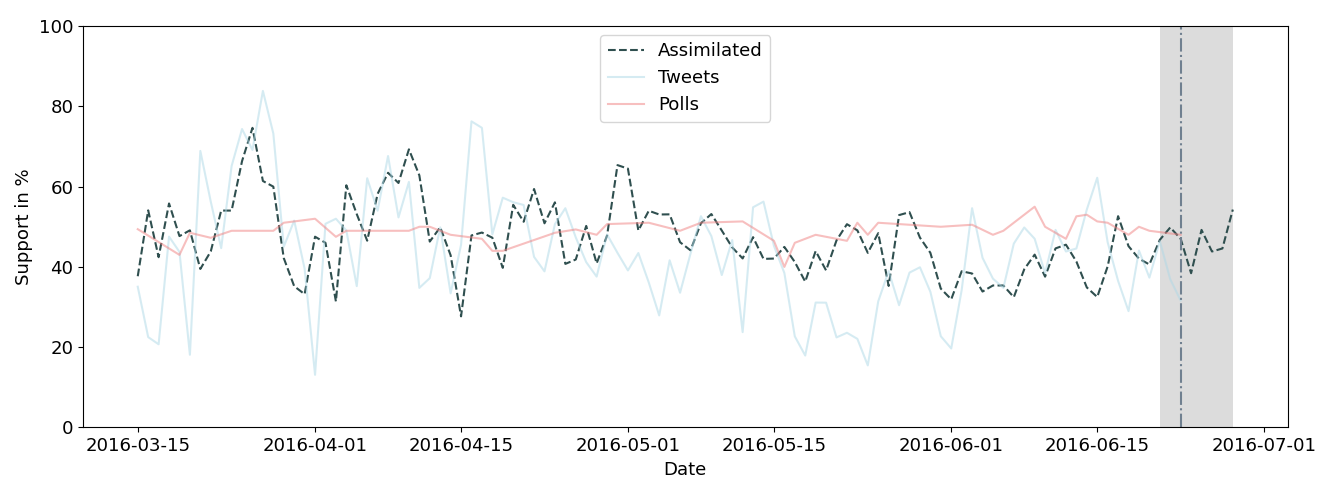} }}%
    \caption{Results for Leave with varying values of $K_k$. Note that the Twitter data plotted is unshifted.}%
    \label{fig:nomodleave}%
\end{figure}

Three different scenarios are discussed: the two edge cases as for the previous OI method, $K_k = 0.43$ and $K_k = 0.95$, and an intermediate value of $K_k = 0.75$. Results are plotted in Figure \ref{fig:nomodremain} for Remain and Figure \ref{fig:nomodleave} for Leave, in which the shaded area only predictions are used since no observations are available. This means that the estimate $x_k^a$ is not computed using the OI state update equation, but simply taken as $x_k$.\newline

The effect of different values of $K_k$ is evident: it clearly dictates the amount of trust placed in either the measurement of the prior estimate, as shown in equation (\ref{stateUp}). At high values of the Kalman gain, the Twitter series barely impacts the assimilated result, and results follow polls closely. For the case of $K_k=0.49$, both series are trusted almost equally. The importance of finding the right K value in this approach can not be understated. This work provides reasonable bounds, and a useful starting point for future research to narrow down the range. The case when $K_k=0.95$ does barely incorporate any information from the Twitter series and closely aligns with the time shifted polls. In the case when $K_k=0.75$ and $K_k=0.49$, however, we see that our results show a strong upward trend in Leave support in the days leading up to the election. In contrast, polling data for Leave measured a decline in support during that period. Our model is hence a more accurate predictor of the election outcome than polls.\newline

Since apart from the election outcome, no ground truth is available, model performance had to be evaluated with non-traditional methods. Firstly, the distance from the assimilated data to the polling data and shifted Twitter data was measured, to establish to which degree information from the two time series propagates into the final result. Results are shown in Table \ref{table:MSEnomodel}. It clearly shows a trade-off between improved Mean Squared Error (MSE) with respect to either polls or Twitter: distance to polling data decreases with increasing values of $K_k$, and distance to Twitter data improves with decreasing values of $K_k$. As expected, higher value of $K_k$ corresponds to closer alignment with polls since the assimilation favours the measurements. The correlation between the assimilated results and the shifted Twitter and polling series was also calculated using equation (\ref{eq:corr}) and results are shown in Table \ref{table:CORRnomodel}. Results for the correlation convey a similar message: lower values of $K_k$ result in a higher correlation with the Twitter data, and higher values for $K_k$ lead to a higher correlation with polls.  \newline

\begin{figure}[h!]
    \centering
{\includegraphics[width=5.6cm]{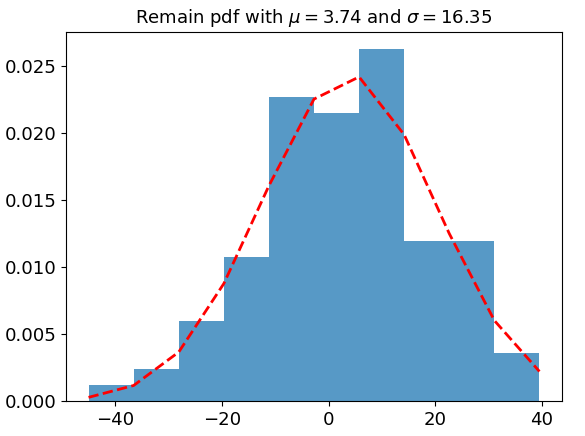} }%
\caption{Results for residual errors for the case of Remain.}%
\label{fig:nomodnorm}%
\end{figure}

\begin{figure}[h!]
    \centering
{\includegraphics[width=5.6cm]{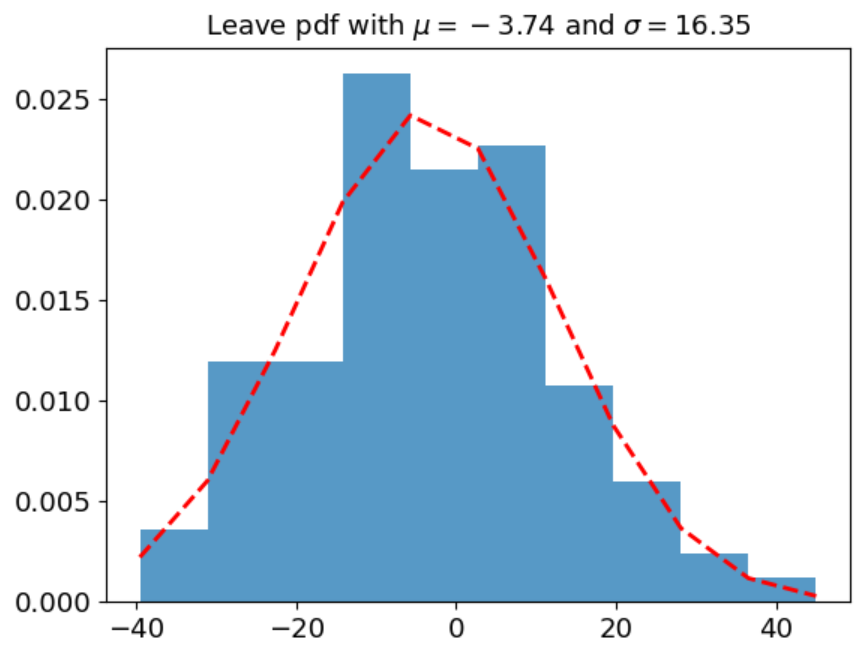} }%
\caption{Results for residual errors for the case of Leave.}%
\label{fig:nomodnormleave}%
\end{figure}

As a final evaluation method, it was checked whether the difference between  $x_k$ and $z_k$ are Gaussian, as required by one of the key OI assumptions. Since the values of $x_k$ are independent from the previously assimilated values $x^a_{k-1}$ in our model, residuals are the same for all three values of $K_k$. Figures \ref{fig:nomodnorm}-\ref{fig:nomodnormleave} show the distribution of residuals. Firstly, results show that errors are approximately Gaussian distributed. It is clear, however, that they are not zero-mean but rather favour Remain. When errors are not centred on zero, The OI update equation is not the optimal estimator, but is still able to perform adequately given the deviations are within an acceptable bound \cite{ensemblekfbook}.\newline

Another important characteristic of our model is its computational efficiency. The computation time for our model averaged 0.177s over 100 different runs. The Twitter classification process, which classified Tweets as supporting Remain or Leave by Amador et al. \cite{lopez2017predicting}, took under 35 hours for the entire batch. Considering this, the computation time of our model is negligible, whilst the results significantly improve upon standalone Twitter or polling data estimates.

\begin{table}[h!]
\centering
\resizebox{!}{!}{%
\begin{tabular}{ccccc}
\hline
\textbf{Value of $K_k$} & \textbf{Polls} & \textbf{Twitter} & \textbf{$R_k\,[\%^2]$} & \textbf{$P^b\,[\%^2]$} \\ \hline
  $0.95$                 & 6.05 & 253.89  & $13.08$    &  $339.52$   \\
 $0.75$                 & 22.66 & 269.24  & $13.08$    &  $45$        \\
 $0.43$                 & 96.05 & 52.02  & $13.08$    & $10$         
     \\ \hline
\end{tabular}%
}
\caption{MSE of the assimilated series with respect to the polling and shifted Twitter series for Remain for different values of $P^b$ and $R_k$.\label{table:MSEnomodel}}
\end{table}

\begin{table}[h!]
\centering
\resizebox{!}{!}{%
\begin{tabular}{ccc}
\hline
\textbf{Value of $K_k$} & \textbf{Polls} & \textbf{Twitter}  \\ \hline
  $0.95$                 & 0.93 & 0.30  \\
 $0.75$                 & 0.32 & 0.92        \\
 $0.43$                 & 0.06 & 0.99          
     \\ \hline
\end{tabular}%
}
\caption{Correlation of the assimilated series with respect to the shifted polling and shifted Twitter series.\label{table:CORRnomodel}}
\end{table}

\section{Conclusion and future work}

This work aimed to correct public opinion estimates by merging both Twitter opinion and polling data through Bayesian data assimilation. It successfully described current challenges and techniques with measuring public opinion through Twitter, proposed an efficient assimilation architecture, and evaluated its performance for the Brexit referendum. Our results clearly showed a strong upward trend in Leave support around the date of the referendum, something that both polling data and Twitter data failed to do when evaluated separately. The model hence improved significantly upon standalone estimates, and did so at a low computational cost. The average computation time comprised 0.177s for the entire data set. One crucial finding to the implementation of our model was the discovery of a time gap between the Twitter and polling data. For our specific data sets, it was found that Twitter opinion preceded public opinion by 14 days, and that polling data lagged by 2 days. Our work also demonstrated the importance of accurate hyperparameter estimation, and provided a useful starting point by providing reasonable bounds for the model parameters. \\
Further research can expand upon this work in a number of ways. Firstly, attempts could be made at estimating a dynamic model of the system, potentially through machine learning methods. This way, more advanced Bayesian data assimilation algorithms can be used, such as the ensemble Kalman Filter. Efforts should also focus on refining model parameter bounds and assessing whether our approach proves succesful in other election scenarios.\newline

\newpage

\printbibliography

\end{document}